\documentclass[preprint2,tighten,times]{aastex6}

\newcommand{\lya}{Ly${\alpha}$}
\newcommand{\clustername}{SDSS~J2222+2745}
\newcommand{\sgasname}{SDSS~J2222+2745}
\newcommand{\hst}{{\it HST}}

\newcommand{\kms}{km s$^{-1}$}

\shortauthors{Bayliss et al.}

\begin{document}

\title{Spatially Resolved Patchy Lyman-$\alpha$ Emission 
Within the Central Kiloparsec of a Strongly Lensed Quasar Host Galaxy at \lowercase{z $=$ 2.8}}

\author{Matthew~B.~Bayliss\altaffilmark{1},  
Keren~Sharon\altaffilmark{2},
Ayan~Acharyya\altaffilmark{3}, 
Michael ~D.~Gladders\altaffilmark{4,5}, 
Jane~R.~Rigby\altaffilmark{6},
Fuyan~Bian\altaffilmark{3,7}, 
Rongmon~Bordoloi\altaffilmark{1,8},
Jessie~Runnoe\altaffilmark{2}, 
Hakon~Dahle\altaffilmark{9}, 
Lisa~Kewley\altaffilmark{3}, 
Michael~Florian\altaffilmark{5}, 
Traci~Johnson\altaffilmark{2}, and
Rachel Paterno-Mahler\altaffilmark{2} }

\altaffiltext{1}{Kavli Institute for Astrophysics and Space Research, Massachusetts Institute of Technology, 77 Massachusetts Avenue, Cambridge, MA 02139, USA }
\altaffiltext{2}{Department of Astronomy, University of Michigan, 1085 S. University Ave, Ann Arbor, MI 48109, USA}
\altaffiltext{3}{RSAA, Australian National University, Cotter Road, Weston Creek, ACT 2611, Australia}
\altaffiltext{4}{Kavli Institute for Cosmological Physics, University of Chicago, Chicago, IL 60637, USA}
\altaffiltext{5}{Department of Astronomy and Astrophysics, University of Chicago, Chicago, IL 60637, USA}
\altaffiltext{6}{Astrophysics Science Division, NASA Goddard Space Flight Center, 8800 Greenbelt Rd., Greenbelt, MD 20771, USA}
\altaffiltext{7}{Stromlo Fellow}
\altaffiltext{8}{Hubble Fellow}
\altaffiltext{9}{Institute of Theoretical Astrophysics, University of Oslo, P.O. Box 1029, Blindern, N-0315 Oslo, Norway}

\email{mbayliss@mit.edu}

\begin{abstract}

We report the detection of extended Lyman-$\alpha$ emission from the host galaxy of SDSS~J2222+2745, 
a strongly lensed quasar at $z = 2.8$. Spectroscopic follow-up clearly reveals extended Lyman-$\alpha$ in 
emission between two images of the central active galactic nucleus (AGN). We reconstruct the lensed quasar 
host galaxy in the source plane by applying a strong lens model to \hst\ imaging, and resolve spatial scales 
as small as $\sim$200 parsecs. In the source plane we recover the host galaxy morphology to within a few 
hundred parsecs of the central AGN, and map the extended Lyman-$\alpha$ emission to its physical origin 
on one side of the host galaxy at radii $\sim$0.5--2 kpc from the central AGN. There are clear morphological 
differences between the Lyman-$\alpha$ and rest-frame ultraviolet stellar continuum emission from the quasar 
host galaxy. Furthermore, the relative velocity profiles of quasar Lyman-$\alpha$, host galaxy Lyman-$\alpha$, 
and metal lines in outflowing gas reveal differences in the absorbing material affecting the AGN and host 
galaxy. These data indicate the presence of patchy local intervening gas in front of the central quasar and its 
host galaxy. This interpretation is consistent with the central luminous quasar being obscured across a 
substantial fraction of its surrounding solid angle, resulting in strong anisotropy in the exposure of the host 
galaxy to ionizing radiation from the AGN. This work demonstrates the power of strong lensing-assisted 
studies to probe spatial scales that are currently inaccessible by other means. 

\end{abstract}


\section{Introduction} \label{sec:intro}

All massive galaxies harbor central super-massive black holes that produce strong emission when they 
accrete in-falling matter. This ubiquitous accretion-emission phenomenon is referred to as an AGN 
\citep{netzer15}. Galaxies with AGN that out-shine their host galaxies (``quasars'') are ubiquitous at 
$z\gtrsim1$ \citep[e.g.;][]{paris17}, i.e., during the epoch in 
which the Universe formed most of its stars \citep{madau14}. However, quasars are transient 
phenomena with duty cycles that are difficult to measure 
\citep{martini03,wang06,shen07,shankar10,schmidt2017}. 
AGN also act as a source of feedback, injecting energy both radiatively and mechanically back into 
their immediate environments. Quasars also easily 
dominate the production of ionizing radiation in their local environment, even fluorescing neutral 
hydrogen out beyond the virial radii of the dark matter haloes in which the quasars reside  
\citep[e.g.,][]{christensen06,smith09,cantalupo14,hennawi15,borisova16,fathivavsari16}. 

\begin{figure*}[t]
\epsscale{1.16}
\plotone{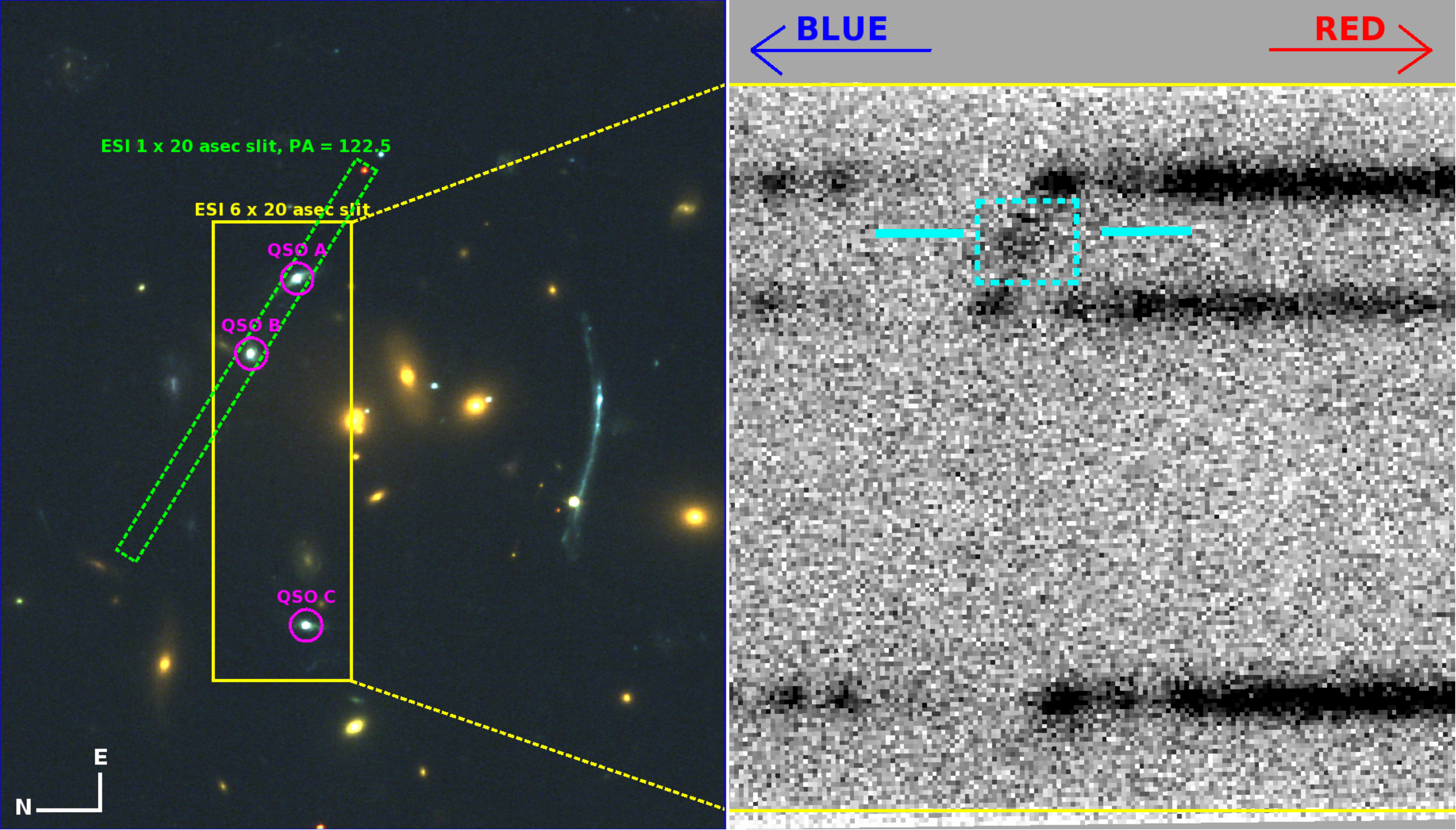}
\caption{{\it Left:} \clustername\ as imaged by \hst\ in bands F435W (blue), F606W (green), and 
F814W (red). The positions of images A, B, and C of the lensed quasar (QSO) are labeled using 
the conventions established in \citet{dahle2013}. The positions of 
both ESI observations using the 1\arcsec$\times$ 20\arcsec\ (green dashed; PA$=122.5$ 
degrees E of N) and the 6\arcsec$\times$ 20\arcsec\ (solid yellow; PA$\sim90$ degrees E of N) 
slits are also shown. 
{\it Right:} 2D ESI spectrum from the 6\arcsec$\times$ 20\arcsec\ ESI slit, zoomed in on 
the region containing \lya\ at the quasar$+$host galaxy redshift ($\lambda \sim 4600$\AA); 
the quasar exhibits a classic asymmetric \lya\ emission profile with strong emission on the 
red side and absorption on the blue side. Shown in 2D, the wavelength solution as a function of 
vertical position along the slit is no constant and depends on the horizontal position of different 
sources within the slit. The host galaxy \lya\ emission is indicated in cyan.}
\label{fig:slitimg}
\end{figure*}

Resolving the central regions of distant quasar host galaxies is challenging. Even subtracting the 
quasar emission with a well-understood point spread function (PSF) in the ultraviolet (UV) 
and optical can leave residual noise from the wings of the quasar emission that overwhelms 
the host galaxy. Recent ALMA observations at millimeter wavelengths---where the central AGN 
emission is less of an issue---can resolve the cold molecular gas content of distant quasar host 
galaxies on $\sim1-2$ kpc scales \citep{wang13,paraficz17,venemans17}. More local, low-redshift 
studies ($z\lesssim0.4$) using adaptive optics and \hst\ can recover host galaxy morphologies in 
the optical and near infrared, and find that quasar hosts can be early or late-type galaxies 
\citep{dunlop03,guyon06} and can exhibit significant star-formation 
\citep[e.g.,][]{young14}. Little is known, however, about how radiation from the quasar directly 
affects atomic gas in their host galaxies, especially within the central kiloparsec during the era of 
peak quasar activity. Currently the only way to probe these regions in the UV/optical is with 
high magnification strong gravitational lensing systems. 

The idea of using lensed quasars to study their host galaxies is not new 
\citep[e.g.;][]{peng06,ross09,oguri13,rusu2014}, and multiply--imaged quasars with large image 
separations are ideal for this purpose. These are 
quasars lensed by groups or clusters of galaxies, with image separations (a proxy for Einstein radius) 
that are $\gtrsim5$\arcsec. The large separation minimizes the confounding effects of the foreground 
lens light, and these systems also tend to have high magnifications over a larger area in the source 
plane, such that the host galaxy is highly magnified out to physical projected radii of several kiloparsecs.
Currently there are only three large separation lensed quasars known: 
SDSS~J1004+4112 \citep{inada03, oguri04, sharon05, fian2016}, 
SDSS~J1029+2623 \citep{inada06,oguri08,oguri13}, and 
SDSS~J2222+2745 \citep{dahle2013,dahle15, sharon17}.
While strong lensing studies of these systems using \hst\ can provide uniquely high spatial resolution, the 
technique has been relatively under-utilized. This is likely attributable to two factors: the small number of 
suitable quasars known, and the in-depth analysis required to fully exploit the strong lensing. There is reason 
for optimism, however, as the number of known systems should increase in light of recent, current, and 
up-coming wide-field optical surveys. The increasing occurrence of strong lensing studies of highly magnified 
systems bodes well. As we develop ever more expertise with lensing-assisted analyses, the tools and techniques 
for this work improve, and the community becomes more familiar with their potential.

In this Letter we analyze Keck/ESI spectroscopy and \hst\ imaging of the quasar, \sgasname, and its 
host galaxy. The high magnification separates emission from the quasar and its host in both \hst\ imaging 
and ground-based spectroscopy. In $\S$~\ref{sec:obs} we describe the data, 
in $\S$~\ref{sec:ana} we measure the relative spatial and velocity structure of \lya\ in the 
quasar and host galaxy. We propose a physical interpretation in $\S$~\ref{sec:disc} and summarize our 
results in  $\S$~\ref{sec:conc}. We assume a flat $\Lambda$CDM cosmology with $\Omega_{M} = 0.3$, 
and $H_{0}=70$ \kms\ Mpc$^{-1}$; in this cosmology 1\arcsec\ corresponds to a physical scale of 7.849 
kpc at the quasar redshift of $z=2.8054$. 

\begin{figure}[ht!]
\epsscale{1.17}
\plotone{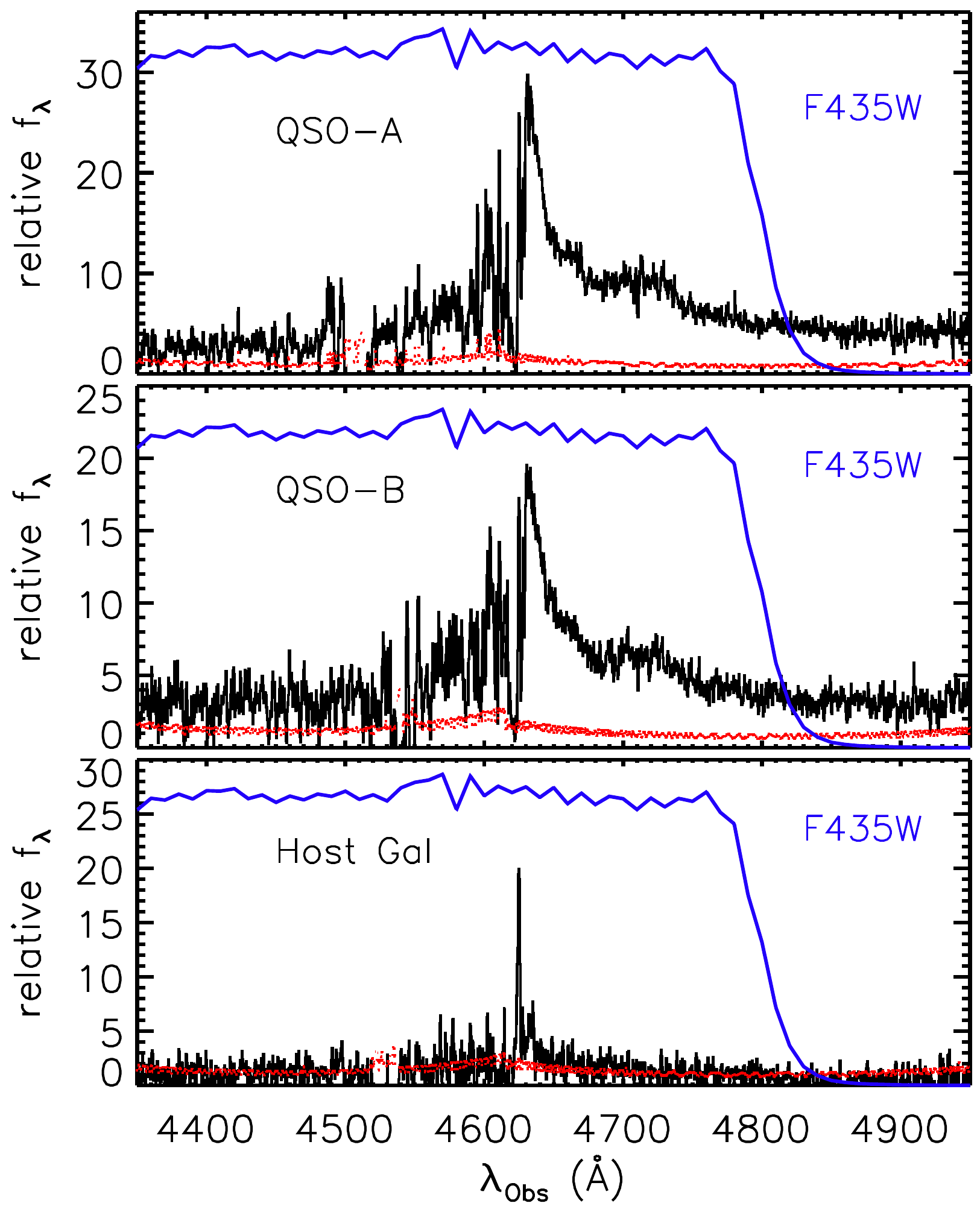}
\caption{
ESI flux (black) with error array (red) of the quasar images A \& B and the host galaxy 
emission between $\Delta \lambda = 4350-4950$\AA, which includes \lya\ ($W_{0} > 20$\AA) plus 
rest-frame UV continuum. The host galaxy spectrum is extracted from the extended emission 
between the spectral traces of quasar images A \& B, as indicated in Figure~\ref{fig:slitimg}. 
The transmission curve of the F435W filter is over-plotted in purple.}
\label{fig:spectra}
\end{figure}

\section{Observations and Data}\label{sec:obs}
\label{sec:obs}

\subsection{\sgasname}

The lensed quasar---SDSS J2222+2745---was discovered in the Sloan Giant Arcs Survey 
\citep[SGAS;][]{koester2010,bayliss2010,bayliss2011a,bayliss2011b,bayliss2014b,sharon14}, 
a systematic search for highly magnified galaxies in the Sloan Digital Sky Survey \citep{york2000}. 
Previously papers described the discovery and initial follow-up \citep{dahle2013}, 
time delay measurements of the three brightest quasar images \citep{dahle15}, and detailed strong 
lens modeling \citep{sharon17} of \sgasname. Here we build on previous studies of this source 
using the lens model from \citet{sharon17}. \sgasname\ is at $z=2.8$, with a luminosity that has 
recently varied between L$_{UV} \simeq 2-5 \times 10^{44}$ erg s$^{-1}$ \citep{dahle15}.

\subsection{Keck/ESI Spectroscopy}

We observed the lensed quasar and its host galaxy with the Echellette Spectrograph and Imager 
(ESI) on the Keck-II telescope on the nights of August 26-27, 2016 (UT dates 2016-08-27 \& 
2016-08-28). On the first night we targeted quasar images A and B with the 1\arcsec\ wide ESI slit 
at a position angle (PA) of 122.5 degrees East of North (Figure~\ref{fig:slitimg}, dashed green), 
completing two 1800 s integrations. The detector was binned by two in the x-direction, i.e. the 
spatial axis along the slit. 

Immediate inspection of the initial 2D spectrum revealed extended \lya\ 
emission between the quasar A and B images (Figure~\ref{fig:slitimg}, right panel). On the second 
night we observed the system again with the 6\arcsec\ wide ESI slit at a PA of 89 degrees East of 
North, simultaneously capturing quasar images A, B and C, as well as the space in between these 
images (Figure~\ref{fig:slitimg}, solid yellow). Our goal was to apply a ``spectral imaging'' mode to 
the entire region where the lensed quasar$+$host galaxy might appear. This second slit position 
revealed extended \lya\ emission between images A and B, but nothing between images B and C 
(Figure~\ref{fig:slitimg}, right panel).

We reduced the ESI spectra using a combination of IRAF tasks and the XIDL ESI pipeline developed 
by J.~X. Prochaska\footnote{http://www2.keck.hawaii.edu/inst/esi/ESIRedux/} to subtract the bias level, 
apply flat-fields, and derive a wavelength map for the data. We used custom IDL routines to subtract a 
model of the sky flux as a function of wavelength, and extract 1-dimensional spectra from several 
manually-defined apertures for spatially distinct emission regions with the lensed quasar$+$host 
galaxy system. All observations were through light cirrus, so we recover only a relative throughput 
calibration as a function of wavelength using observations of two standard stars, BD$+$33 2642 and 
GD50, from the 
CALSPEC\footnote{http://www.stsci.edu/hst/observatory/crds/calspec.html} Calibration Database.  

The final 1\arcsec\ slit 1D spectra have a full width at half max (FWHM) spectral 
resolution, measured from sky lines, of R$=\lambda/\Delta \lambda \simeq 5000$ ($\Delta v \simeq 60$ 
\kms). The seeing was $\sim$0.8\arcsec, smaller than the slit width, and as a result, the spectra of the 
point source quasar images have slightly better spectral resolution (R$ \simeq 5500$) measured for the 
narrowest spectral lines in the data. Unless otherwise indicated, we use spectra extracted from the 
1\arcsec\ slit throughout this analysis due to the significantly lower sky background relative to 
the 6\arcsec\ slit data. 

\subsection{\hst\ Imaging}

We use \hst\ imaging data from Cycle 21 program GO-13337 (PI: Sharon), 
described in detail in \citet{sharon17}. Briefly, these data include {\it WFC3/IR} imaging in F110W and 
F160W, and {\it ACS} imaging in F435W, F606W, and F814W. These bands were selected to 
provide a wide lever-arm in wavelength to optimally constrain the spectral energy distributions of 
sources. The redder filters all sample host galaxy stellar continuum emission at $z=2.8$, while on 
the blue side F435W samples a combination of far-UV stellar continuum, \lya, and the Lyman continuum 
emission at the quasar redshift (Figure~\ref{fig:spectra}). We do not detect the host galaxy continuum 
spectroscopically, implying a rest-frame \lya\ equivalent width, $W_{0} < -20$\AA. The other 
filters sample the UV-through-optical continuum at $z=2.8$: $\lambda_{rest,F606W} \sim 1560$\AA, 
$\lambda_{rest,F814W} \sim 2120$\AA, $\lambda_{rest,F110W } \sim 3030$\AA, and 
$\lambda_{rest,F160W} \sim 4040$\AA. 

\section{Analysis}
\label{sec:ana}

\subsection{Source Plane Reconstruction}

\begin{figure}[h!]
\center
\epsscale{1.14}
\plotone{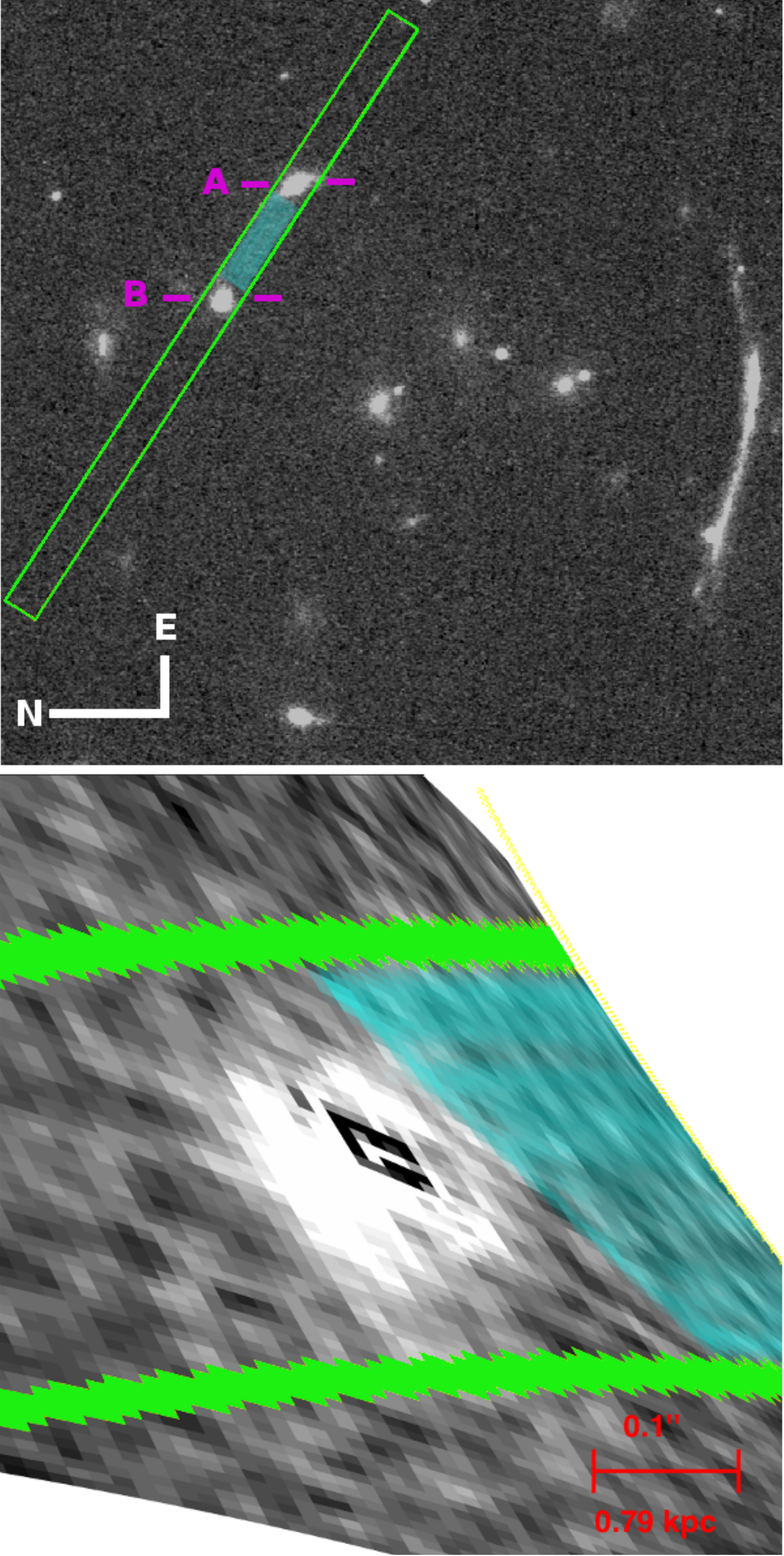}
\caption{
\emph{Top:} F435W image with the 1\arcsec\ ESI slit indicated in yellow, and the region used to 
extract the host galaxy \lya\ marked in cyan (between the traces of lensed quasar images A \& B).
\emph{Bottom:} F435W source plane image of the quasar host galaxy, minus the quasar emission, 
with the 1\arcsec\ ESI slit and host galaxy \lya\ extraction regions from the top panel mapped into 
the source plane. Due to the lensing configuration, the ESI slit samples the fainter side of the host 
galaxy in F435W.}
\label{fig:slitmap}
\end{figure}

We use the strong lensing models and methodology described in \citet{sharon17} to recover 
lensing-corrected source plane images 
of the quasar host galaxy using the most highly magnified quasar image (A). We also map the ESI 
slit into the source plane---shown in Figure~\ref{fig:slitmap}---which reveals the location from which 
the extracted host galaxy \lya\ emission originates. The spatial resolution in the source plane is 
described by an effective lensing PSF. This lensing PSF is elliptical and varies spatially in the 
source plane, with FWHM values of $\sim200 \times 500$ pc in F435W at the location of the 
central AGN (Figure~\ref{fig:srcpln}).

\subsection{Subtraction of the Quasar Light}

\begin{figure*}[ht!]
\center
\epsscale{1.175}
\plotone{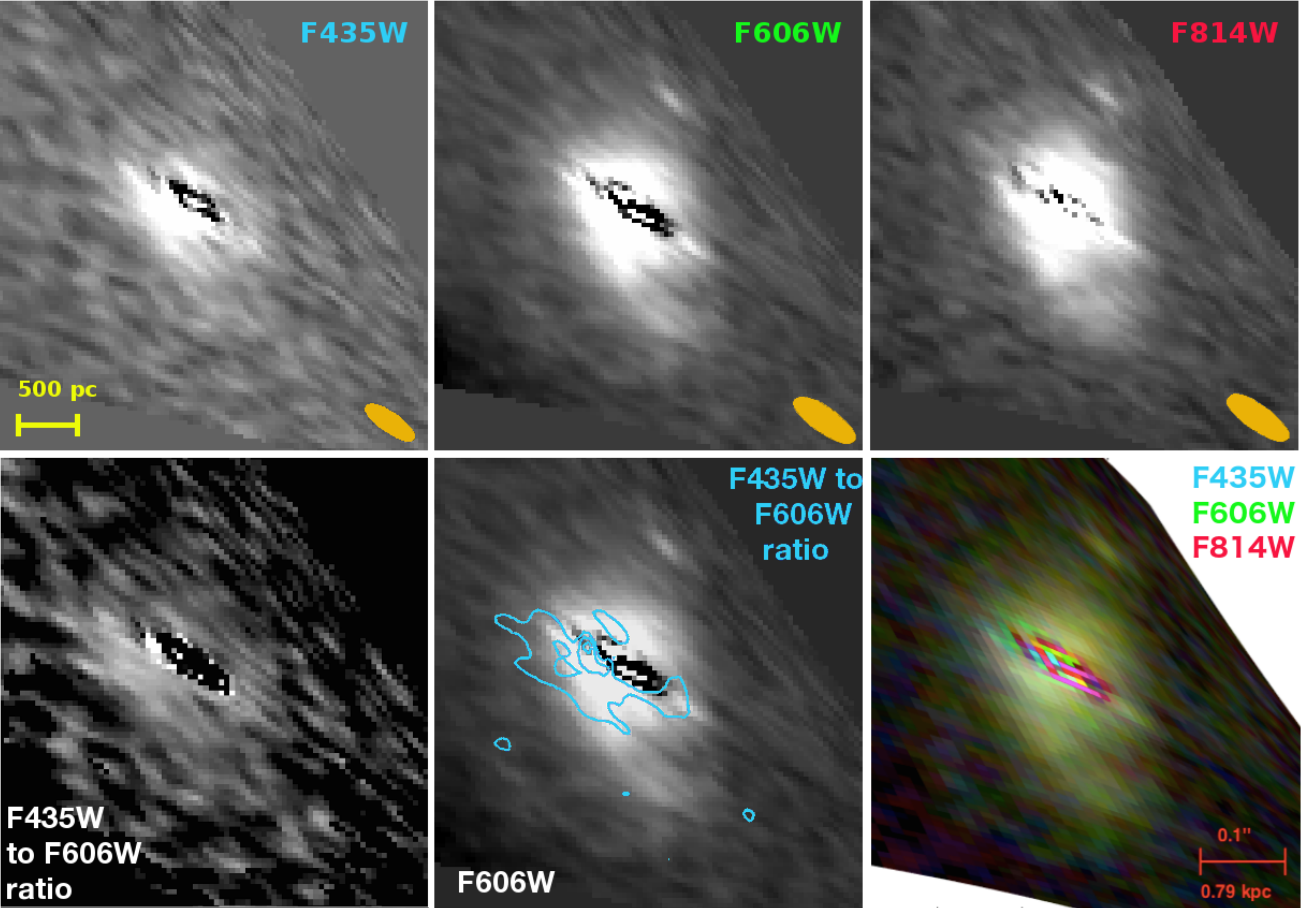}
\caption{
Top three panels: the reconstructed source plane images of the quasar host galaxy after 
subtracting off the AGN flux in F435W, F606W, and F814W. Orange ellipses indicate 
the lensing PSF at the position of the central AGN. The high magnification confines the residuals 
from the quasar subtraction to the central $\lesssim$200-500 parsecs in the bluest 
bands. The bottom three panels show, from right to left, the ratio of F435W--to--F606W flux, 
the F606W image with the contours of the F435W--to--F606W ratio overlaid in cyan, and an 
RGB color image of the host galaxy made from F814W (red), F606W (green), and F435W (blue), 
reproduced from \citet{sharon17}. 
There is a strong asymmetry to the ratio of F435W, which samples \lya, to host galaxy stellar light 
as traced by F606W, and the color of the extended stellar emission is uniform across the entire 
host galaxy.} 
\label{fig:srcpln}
\end{figure*}

We subtract the light from the quasar in the \hst\ images using an isolated star $\sim$20\arcsec\ 
away from the quasar. We scale the peak emission of the star to that of the quasar, subtract the 
result from the lensed quasar images, and use the resulting quasar-subtracted image plane data to 
generate images of the quasar-subtracted host galaxy in the source plane.
We make no attempt to recover spatial information about the host galaxy where the quasar-subtraction 
residuals are large. In F435W this is an elliptical region with major and minor axes of 650 and 275 
parsecs, respectively. The top three panels of Figure~\ref{fig:srcpln} show the host galaxy in the 
source plane in three \hst\ bands that sample \lya\ and the rest-frame UV, in the rest frame. 

\subsection{Systemic Redshift}
\label{sec:systvel}

We measure the systemic redshift using rest-frame UV features; specifically, we take a weighted 
average of the redshifts of O {\small III}] $\lambda$1666\AA\ and C {\small III}] $\lambda\lambda$1907, 
1909\AA\ as measured from gaussian fits to the line profiles. 
We measure a systemic redshift of $z=2.8054 \pm 0.0004$, consistent with our previous measurement 
from $R\sim1000$ spectra \citep{sharon17}. Recent studies show that many quasar emission lines have 
velocities that are biased relative to systemic as measured from stellar features \citep{shen16}, 
suggesting that our systemic redshift estimate uncertainty is dominated by systematics on the order of 
$\delta z \simeq 0.002$. Based on the velocities of other features discussed below, it seems likely that 
the true systemic redshift of \sgasname\ higher than our measurement by $\sim$50-100 \kms.

\begin{figure*}[t!]
\epsscale{1.17}
\plotone{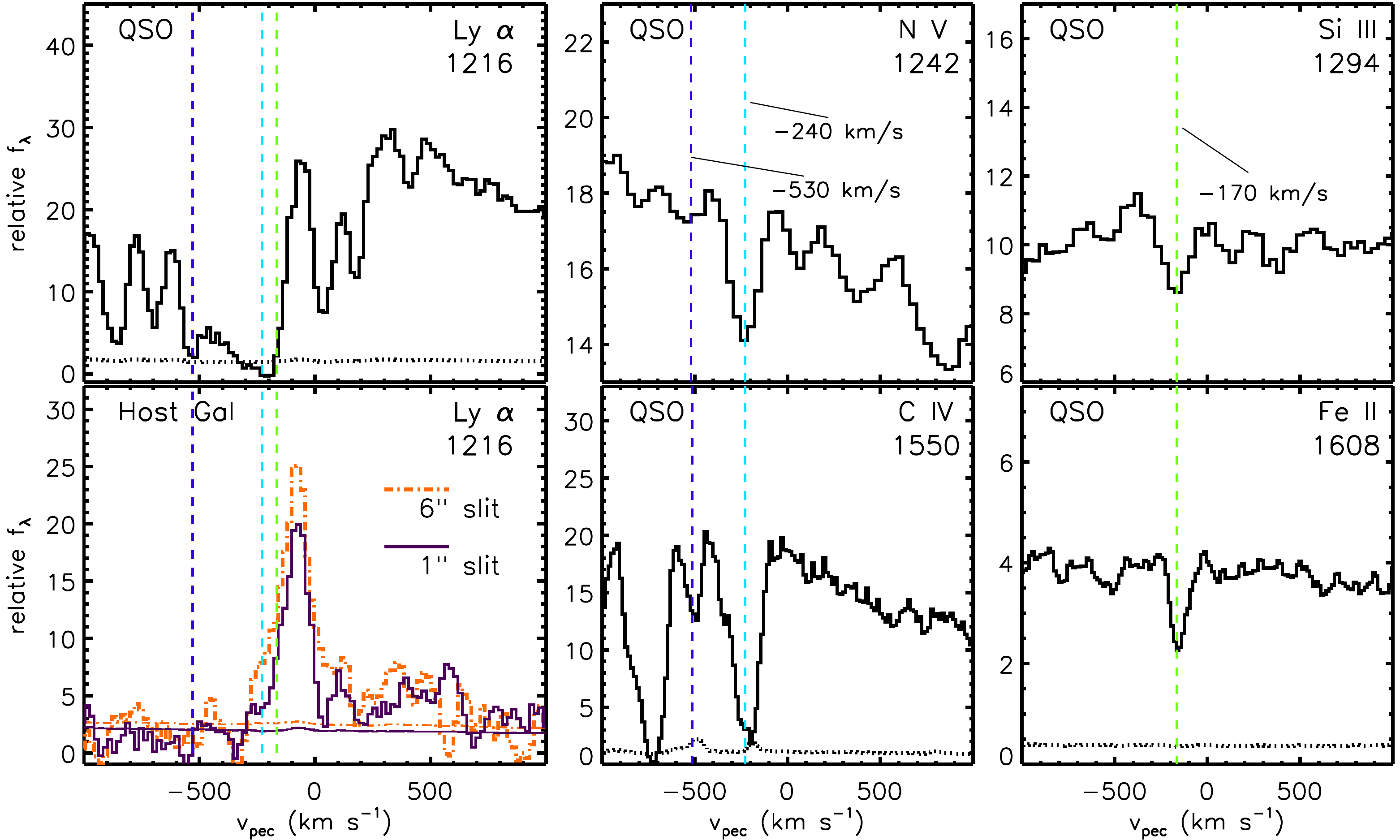}
\caption{
Velocity profiles of \lya\ and metal lines present in the spectrum of lensed quasar \sgasname\ with 
vertical dashed lines indicating outflow velocities traced by metal lines at $-530$ \kms\ (purple), 
$-240$ \kms\ (cyan), and $-170$ (green) \kms. The leftmost two panels show the \lya\ line for the 
quasar (top) and host galaxy (bottom), including the host galaxy spectrum extracted from the 
1\arcsec\ (black solid) and 6\arcsec\ slits (orange dot-dashed). The other panels show the velocity 
profiles of four metal lines that appear in the quasar spectra. 
The blue side of the host galaxy \lya\ emission profile extends to velocities where the quasar \lya\ is strongly 
absorbed, while the red side exhibits weak, choppy emission, likely resulting from either scattered AGN 
\lya, or partially absorbed host galaxy emission.}
\label{fig:velspectra}
\end{figure*}

\subsection{Absorbing Outflows}

The ESI spectra reveal absorption systems as traced by ionized metal-enriched gas. These systems 
have peculiar velocities, $v_{pec}$, that are blueshifted 
relative to systemic, consistent with absorbing, outflowing material. Specifically, we identify  
a low-ionization system---likely associated with a large scale galactic outflow---at $v_{pec} = -170 \pm 
10$ \kms\ from weak Si {\small II} $\lambda$1294 and Fe {\small II} $\lambda$1608 absorption.
We also identify a system at $v_{pec} = -240 \pm 10$ \kms\ that exhibits strong absorption 
by high-ionization lines, including N {\small V} $\lambda$$\lambda$1238, 1242, 
C {\small IV} $\lambda$$\lambda$ 1548, 1550,  Si {\small IV} $\lambda$$\lambda$1393, 1402, and 
He {\small II} $\lambda$1640. A second, much weaker system at $v_{pec} = -540 \pm 10$ \kms\ also 
appears in N {\small V}, C {\small IV}, and Si {\small IV}. The presence of high ionization metals and 
larger outflow velocities suggest that these two system are associated with gas that is more local to 
the central AGN than the lower-ionization, system at $v_{pec} = -170$ \kms. We show the velocity 
profiles of \lya\ and metal lines in Figure~\ref{fig:velspectra}. 

\subsection{ \lya\ Emission}

Figure~\ref{fig:slitmap} shows that our spectroscopic slit samples a region that extends radially away 
from the central AGN by $\sim0.5-2$ kpc where the host galaxy has relatively low surface brightness 
in F435W. The larger collecting area and higher spectral 
resolution results in the ESI spectrum being much more sensitive to the \lya\ emission line than the F435W 
imaging, with respective \lya\ line surface brightness depths of $\sim 2 \times 10^{-19}$ erg s$^{-1}$ cm$^{-1}$ 
arcsec$^{-2}$ (ESI) vs. $\sim 5 \times 10^{-18}$ erg s$^{-1}$ cm$^{-1}$ arcsec$^{-2}$ (F435W). The 
spectroscopic detection confirms relatively low-level \lya\ emission across the entire host galaxy, 
while the F435W image reveals a strong asymmetry in the host galaxy \lya\ emission 
(Figure~\ref{fig:srcpln}, top left panel).

The strongest host galaxy \lya\ emission in the ESI spectrum is $\sim$200 \kms\ wide with a velocity centroid 
that coincides with a peak in quasar emission at $v_{pec} = -80 \pm 10$ \kms, though the host galaxy emission 
extends to bluer velocities than the quasar (Figure~\ref{fig:velspectra}). The quasar 
emission is broad, peaking at velocities of $\sim200-1200$ \kms, where there is only extremely weak host 
galaxy emission. This weak host galaxy \lya\ emission has a choppy velocity structure that is unlike the 
redshifted \lya\ emission profiles seen in starburst galaxies \citep[e.g.,][]{tapken2007,u2015}. We verify 
that this weak emission does indeed originate from the host galaxy by experimenting with several extraction 
regions for host galaxy emission that are increasingly well-separated from the A \& B quasar image traces 
along the slit, i.e., shrinking the cyan region in Figure~\ref{fig:slitmap}. The extended emission at $100 \lesssim 
v_{pec} \lesssim 500$ \kms\ is present independent of how conservatively we define the extraction. 

\section{Physical Interpretation and Discussion}
\label{sec:disc}

The variable F435W--to--F606W flux ratio across the host galaxy implies that \lya\ emission is either differently 
absorbed or emitted along different lines of sight toward the \sgasname\ host galaxy. F606W and F814W 
sample the rest-UV stellar continuum, i.e., light associated with massive blue stars. The rest-UV color of 
the stellar emission is similar across the host galaxy (Figure~\ref{fig:srcpln}, bottom rightmost panel), 
implying that the properties of the massive stars are similar. This implies that the F606W stellar flux should 
map consistently to an ionizing photon flux, which would imply a relatively constant F435W/F606W ratio 
for \lya\ emission that is powered by massive stars alone.

The F435W/F606W emission ratio in the bottom left panel of Figure~\ref{fig:srcpln} shows that the 
spatial distribution of host 
galaxy \lya\ emission is clearly offset relative to the stellar continuum emission. This strong asymmetry 
in F435W/F606W and its offset relative to the stellar continuum (F606W) argue for spatial variations in 
one or both of: 1) the \lya\ emitting gas, possibly from anisotropic illuminating ionizing radiation by the 
central AGN, and 2) the covering factor of intervening neutral hydrogen available to absorb \lya. It is 
possible for these two effects to be related, as asymmetric escaping AGN emission could efficiently 
photo-ionize hydrogen along some sight-lines into the surrounding host galaxy, while leaving significant 
neutral hydrogen along others. Escaping \lya\ emission centered at $v_{pec} = -80$ \kms\ appears in 
both the quasar and host galaxy spectra, consistent with an approximately isotropic bubble of ionized 
hydrogen at a velocity near systemic. The relative velocity structures of \lya\ and the outflows traced 
by metal absorption, shown in Figure~\ref{fig:velspectra}, reveal that the strongest outflow system 
($v_{pec} = -240$ \kms, cyan in Figure~\ref{fig:velspectra}) is coincident with damped \lya\ absorption 
in the quasar spectrum, while the host galaxy \lya\ emission is nonzero at the same velocity. This 
clear spatial de-coupling of windows of \lya\ escape is further evidence of differential absorption in 
front of the AGN versus the host galaxy.

Analysis of hundreds of luminous quasars indicates that they are typically obscured over approximately 
$\sim60$\% of their surrounding solid angle \citep{polletta08}. That kind of obscuration is fully consistent 
with the observed asymmetry in \lya\ inferred from the F435W-to--F606W ratio in \sgasname. It also 
makes sense in light of the weak host galaxy \lya\ emission at $v_{pec} \sim 100-500$ \kms. This 
emission must result either from AGN light being scattered off of dust and gas before escaping, or from 
spatially extended 
\lya\ emission from the host galaxy that is partially covered by scattering/absorbing material. Both 
scenarios imply the presence of more intervening gas and dust in front of the extended host galaxy 
than the AGN, in agreement with the physical picture outlined above in which a substantial fraction of the 
solid angle around the AGN is obscured. 

In this interpretation the quasar can photo-ionize gas along some lines of sight out of its host galaxy, but is 
obscured along others. The source of the obscuring material should be some combination of a dusty torus 
or gas and dust on large scales \citep{dipompeo16}. The morphology that we observe here within the 
central few kiloparsecs is similar to previously observed asymmetry on $\sim$100's of kpc scales 
in \lya\ haloes around distant quasars \citep{cantalupo14,hennawi15,borisova16}, which is also consistent 
with central AGN anisotropically illuminating their surrounding environments with ionizing photons.

\section{Summary}
\label{sec:conc}

Strong gravitational lensing of \sgasname\ enables a source reconstruction of the rest-frame UV emission 
in the host galaxy core of this distant quasar. The ESI spectroscopy and \hst\ imaging combine to reveal 
direct evidence of patchy structure in the \lya\ emission from the host galaxy of a $z=2.8$ quasar. We 
attribute this structure to anisotropic ionization and/or illumination of gas. \sgasname\ is an ideal 
candidate for additional follow-up to study the properties of a quasar host galaxy on sub-kpc scales. 
Future integral field unit (IFU) spectroscopy with $JWST$ and narrow-band \hst\ imaging of this and 
other similar systems would enable detailed emission line studies of the relative spatial and velocity 
structure of gas in quasar host galaxies.

\acknowledgments

This work was supported in part by a NASA Keck PI Data Award, administered by the NASA 
Exoplanet Science Institute, by NASA grant HST-GO-13337, and by an Australian Government 
astronomy research infrastructure grant, via the Department of Industry and Science. RB was 
supported by NASA through Hubble Fellowship grant \#51354 awarded by the Space Telescope 
Science Institute, which is operated by the Association of Universities for Research in Astronomy, 
Inc., for NASA, under contract NAS 5-26555.
Data presented herein were obtained at the W. M. Keck Observatory from telescope time allocated to 
the National Aeronautics and Space Administration through the agency's scientific partnership with the 
California Institute of Technology and the University of California, and also to the Australian community 
through the Australian National Collaborative Research Infrastructure Strategy, via the Department of 
Education and Training. The Observatory was made possible by the generous financial support of the 
W. M. Keck Foundation. The authors wish to recognize and acknowledge the very significant cultural 
role and reverence that the summit of Mauna Kea has always had within the indigenous Hawaiian 
community. Finally, we thank the anonymous referee for thoughtful feedback that improved the 
quality of this paper.

\facilities{HST(ACS,WFC3), Keck:II(ESI)}

\bibliographystyle{aasjournal}

\begin{thebibliography}{}
\expandafter\ifx\csname natexlab\endcsname\relax\def\natexlab#1{#1}\fi

\bibitem[{{Bayliss} {et~al.}(2011{\natexlab{a}}){Bayliss}, {Gladders}, {Oguri},
  {Hennawi}, {Sharon}, {Koester}, \& {Dahle}}]{bayliss2011a}
{Bayliss}, M.~B., {Gladders}, M.~D., {Oguri}, M., {et~al.} 2011{\natexlab{a}},
  \apjl, 727, L26+

\bibitem[{{Bayliss} {et~al.}(2011{\natexlab{b}}){Bayliss}, {Hennawi},
  {Gladders}, {Koester}, {Sharon}, {Dahle}, \& {Oguri}}]{bayliss2011b}
{Bayliss}, M.~B., {Hennawi}, J.~F., {Gladders}, M.~D., {et~al.}
  2011{\natexlab{b}}, \apjs, 193, 8

\bibitem[{{Bayliss} {et~al.}(2014){Bayliss}, {Rigby}, {Sharon}, {Wuyts},
  {Florian}, {Gladders}, {Johnson}, \& {Oguri}}]{bayliss2014b}
{Bayliss}, M.~B., {Rigby}, J.~R., {Sharon}, K., {et~al.} 2014, \apj, 790, 144

\bibitem[{{Bayliss} {et~al.}(2010){Bayliss}, {Wuyts}, {Sharon}, {Gladders},
  {Hennawi}, {Koester}, \& {Dahle}}]{bayliss2010}
{Bayliss}, M.~B., {Wuyts}, E., {Sharon}, K., {et~al.} 2010, \apj, 720, 1559

\bibitem[{{Borisova} {et~al.}(2016){Borisova}, {Cantalupo}, {Lilly}, {Marino},
  {Gallego}, {Bacon}, {Blaizot}, {Bouch{\'e}}, {Brinchmann}, {Carollo},
  {Caruana}, {Finley}, {Herenz}, {Richard}, {Schaye}, {Straka}, {Turner},
  {Urrutia}, {Verhamme}, \& {Wisotzki}}]{borisova16}
{Borisova}, E., {Cantalupo}, S., {Lilly}, S.~J., {et~al.} 2016, \apj, 831, 39

\bibitem[{{Cantalupo} {et~al.}(2014){Cantalupo}, {Arrigoni-Battaia},
  {Prochaska}, {Hennawi}, \& {Madau}}]{cantalupo14}
{Cantalupo}, S., {Arrigoni-Battaia}, F., {Prochaska}, J.~X., {Hennawi}, J.~F.,
  \& {Madau}, P. 2014, \nat, 506, 63

\bibitem[{{Christensen} {et~al.}(2006){Christensen}, {Jahnke}, {Wisotzki}, \&
  {S{\'a}nchez}}]{christensen06}
{Christensen}, L., {Jahnke}, K., {Wisotzki}, L., \& {S{\'a}nchez}, S.~F. 2006,
  \aap, 459, 717

\bibitem[{{Dahle} {et~al.}(2015){Dahle}, {Gladders}, {Sharon}, {Bayliss}, \&
  {Rigby}}]{dahle15}
{Dahle}, H., {Gladders}, M.~D., {Sharon}, K., {Bayliss}, M.~B., \& {Rigby},
  J.~R. 2015, \apj, 813, 67

\bibitem[{{Dahle} {et~al.}(2013){Dahle}, {Gladders}, {Sharon}, {Bayliss},
  {Wuyts}, {Abramson}, {Koester}, {Groeneboom}, {Brinckmann}, {Kristensen},
  {Lindholmer}, {Nielsen}, {Krogager}, \& {Fynbo}}]{dahle2013}
{Dahle}, H., {Gladders}, M.~D., {Sharon}, K., {et~al.} 2013, \apj, 773, 146

\bibitem[{{DiPompeo} {et~al.}(2016){DiPompeo}, {Runnoe}, {Hickox}, {Myers}, \&
  {Geach}}]{dipompeo16}
{DiPompeo}, M.~A., {Runnoe}, J.~C., {Hickox}, R.~C., {Myers}, A.~D., \&
  {Geach}, J.~E. 2016, \mnras, 460, 175

\bibitem[{{Dunlop} {et~al.}(2003){Dunlop}, {McLure}, {Kukula}, {Baum}, {O'Dea},
  \& {Hughes}}]{dunlop03}
{Dunlop}, J.~S., {McLure}, R.~J., {Kukula}, M.~J., {et~al.} 2003, \mnras, 340,
  1095

\bibitem[{{Fathivavsari} {et~al.}(2016){Fathivavsari}, {Petitjean},
  {Noterdaeme}, {P{\^a}ris}, {Finley}, {L{\'o}pez}, \&
  {Srianand}}]{fathivavsari16}
{Fathivavsari}, H., {Petitjean}, P., {Noterdaeme}, P., {et~al.} 2016, \mnras,
  461, 1816

\bibitem[{{Fian} {et~al.}(2016){Fian}, {Mediavilla}, {Hanslmeier}, {Oscoz},
  {Serra-Ricart}, {Mu{\~n}oz}, \& {Jim{\'e}nez-Vicente}}]{fian2016}
{Fian}, C., {Mediavilla}, E., {Hanslmeier}, A., {et~al.} 2016, \apj, 830, 149

\bibitem[{{Guyon} {et~al.}(2006){Guyon}, {Sanders}, \& {Stockton}}]{guyon06}
{Guyon}, O., {Sanders}, D.~B., \& {Stockton}, A. 2006, \apjs, 166, 89

\bibitem[{{Hennawi} {et~al.}(2015){Hennawi}, {Prochaska}, {Cantalupo}, \&
  {Arrigoni-Battaia}}]{hennawi15}
{Hennawi}, J.~F., {Prochaska}, J.~X., {Cantalupo}, S., \& {Arrigoni-Battaia},
  F. 2015, Science, 348, 779

\bibitem[{{Inada} {et~al.}(2003){Inada}, {Oguri}, {Pindor}, {Hennawi}, {Chiu},
  {Zheng}, {Ichikawa}, {Gregg}, {Becker}, {Suto}, {Strauss}, {Turner},
  {Keeton}, {Annis}, {Castander}, {Eisenstein}, {Frieman}, {Fukugita}, {Gunn},
  {Johnston}, {Kent}, {Nichol}, {Richards}, {Rix}, {Sheldon}, {Bahcall},
  {Brinkmann}, {Ivezi{\'c}}, {Lamb}, {McKay}, {Schneider}, \& {York}}]{inada03}
{Inada}, N., {Oguri}, M., {Pindor}, B., {et~al.} 2003, \nat, 426, 810

\bibitem[{{Inada} {et~al.}(2006){Inada}, {Oguri}, {Morokuma}, {Doi}, {Yasuda},
  {Becker}, {Richards}, {Kochanek}, {Kayo}, {Konishi}, {Utsunomiya}, {Shin},
  {Strauss}, {Sheldon}, {York}, {Hennawi}, {Schneider}, {Dai}, \&
  {Fukugita}}]{inada06}
{Inada}, N., {Oguri}, M., {Morokuma}, T., {et~al.} 2006, \apjl, 653, L97

\bibitem[{{Koester} {et~al.}(2010){Koester}, {Gladders}, {Hennawi}, {Sharon},
  {Wuyts}, {Rigby}, {Bayliss}, \& {Dahle}}]{koester2010}
{Koester}, B.~P., {Gladders}, M.~D., {Hennawi}, J.~F., {et~al.} 2010, \apjl,
  723, L73

\bibitem[{{Madau} \& {Dickinson}(2014)}]{madau14}
{Madau}, P., \& {Dickinson}, M. 2014, \araa, 52, 415

\bibitem[{{Martini} \& {Schneider}(2003)}]{martini03}
{Martini}, P., \& {Schneider}, D.~P. 2003, \apjl, 597, L109

\bibitem[{{Netzer}(2015)}]{netzer15}
{Netzer}, H. 2015, \araa, 53, 365

\bibitem[{{Oguri} {et~al.}(2004){Oguri}, {Inada}, {Keeton}, {Pindor},
  {Hennawi}, {Gregg}, {Becker}, {Chiu}, {Zheng}, {Ichikawa}, {Suto}, {Turner},
  {Annis}, {Bahcall}, {Brinkmann}, {Castander}, {Eisenstein}, {Frieman},
  {Goto}, {Gunn}, {Johnston}, {Kent}, {Nichol}, {Richards}, {Rix}, {Schneider},
  {Sheldon}, \& {Szalay}}]{oguri04}
{Oguri}, M., {Inada}, N., {Keeton}, C.~R., {et~al.} 2004, \apj, 605, 78

\bibitem[{{Oguri} {et~al.}(2008){Oguri}, {Ofek}, {Inada}, {Morokuma}, {Falco},
  {Kochanek}, {Kayo}, {Broadhurst}, \& {Richards}}]{oguri08}
{Oguri}, M., {Ofek}, E.~O., {Inada}, N., {et~al.} 2008, \apjl, 676, L1

\bibitem[{{Oguri} {et~al.}(2013){Oguri}, {Schrabback}, {Jullo}, {Ota},
  {Kochanek}, {Dai}, {Ofek}, {Richards}, {Blandford}, {Falco}, \&
  {Fohlmeister}}]{oguri13}
{Oguri}, M., {Schrabback}, T., {Jullo}, E., {et~al.} 2013, \mnras, 429, 482

\bibitem[{{Paraficz} {et~al.}(2017){Paraficz}, {Rybak}, {McKean}, {Vegetti},
  {Sluse}, {Courbin}, {Stacey}, {Suyu}, {Dessauges-Zavadsky}, {Fassnacht}, \&
  {Koopmans}}]{paraficz17}
{Paraficz}, D., {Rybak}, M., {McKean}, J.~P., {et~al.} 2017, ArXiv e-prints,
  arXiv:1705.09931

\bibitem[{{P{\^a}ris} {et~al.}(2017){P{\^a}ris}, {Petitjean}, {Ross}, {Myers},
  {Aubourg}, {Streblyanska}, {Bailey}, {Armengaud}, {Palanque-Delabrouille},
  {Y{\`e}che}, {Hamann}, {Strauss}, {Albareti}, {Bovy}, {Bizyaev}, {Niel
  Brandt}, {Brusa}, {Buchner}, {Comparat}, {Croft}, {Dwelly}, {Fan},
  {Font-Ribera}, {Ge}, {Georgakakis}, {Hall}, {Jiang}, {Kinemuchi},
  {Malanushenko}, {Malanushenko}, {McMahon}, {Menzel}, {Merloni}, {Nandra},
  {Noterdaeme}, {Oravetz}, {Pan}, {Pieri}, {Prada}, {Salvato}, {Schlegel},
  {Schneider}, {Simmons}, {Viel}, {Weinberg}, \& {Zhu}}]{paris17}
{P{\^a}ris}, I., {Petitjean}, P., {Ross}, N.~P., {et~al.} 2017, \aap, 597, A79

\bibitem[{{Peng} {et~al.}(2006){Peng}, {Impey}, {Rix}, {Kochanek}, {Keeton},
  {Falco}, {Leh{\'a}r}, \& {McLeod}}]{peng06}
{Peng}, C.~Y., {Impey}, C.~D., {Rix}, H.-W., {et~al.} 2006, \apj, 649, 616

\bibitem[{{Polletta} {et~al.}(2008){Polletta}, {Weedman}, {H{\"o}nig},
  {Lonsdale}, {Smith}, \& {Houck}}]{polletta08}
{Polletta}, M., {Weedman}, D., {H{\"o}nig}, S., {et~al.} 2008, \apj, 675, 960

\bibitem[{{Ross} {et~al.}(2009){Ross}, {Assef}, {Kochanek}, {Falco}, \&
  {Poindexter}}]{ross09}
{Ross}, N.~R., {Assef}, R.~J., {Kochanek}, C.~S., {Falco}, E., \& {Poindexter},
  S.~D. 2009, \apj, 702, 472

\bibitem[{{Rusu} {et~al.}(2014){Rusu}, {Oguri}, {Minowa}, {Iye}, {More},
  {Inada}, \& {Oya}}]{rusu2014}
{Rusu}, C.~E., {Oguri}, M., {Minowa}, Y., {et~al.} 2014, \mnras, 444, 2561

\bibitem[{{Schmidt} {et~al.}(2017){Schmidt}, {Worseck}, {Hennawi}, {Prochaska},
  \& {Crighton}}]{schmidt2017}
{Schmidt}, T.~M., {Worseck}, G., {Hennawi}, J.~F., {Prochaska}, J.~X., \&
  {Crighton}, N.~H.~M. 2017, ArXiv e-prints, arXiv:1701.08769

\bibitem[{{Shankar} {et~al.}(2010){Shankar}, {Weinberg}, \& {Shen}}]{shankar10}
{Shankar}, F., {Weinberg}, D.~H., \& {Shen}, Y. 2010, \mnras, 406, 1959

\bibitem[{{Sharon} {et~al.}(2014){Sharon}, {Gladders}, {Rigby}, {Wuyts},
  {Bayliss}, {Johnson}, {Florian}, \& {Dahle}}]{sharon14}
{Sharon}, K., {Gladders}, M.~D., {Rigby}, J.~R., {et~al.} 2014, \apj, 795, 50

\bibitem[{{Sharon} {et~al.}(2005){Sharon}, {Ofek}, {Smith}, {Broadhurst},
  {Maoz}, {Kochanek}, {Oguri}, {Suto}, {Inada}, \& {Falco}}]{sharon05}
{Sharon}, K., {Ofek}, E.~O., {Smith}, G.~P., {et~al.} 2005, \apjl, 629, L73

\bibitem[{{Sharon} {et~al.}(2017){Sharon}, {Bayliss}, {Dahle}, {Florian},
  {Gladders}, {Johnson}, {Paterno-Mahler}, {Rigby}, {Whitaker}, \&
  {Wuyts}}]{sharon17}
{Sharon}, K., {Bayliss}, M.~B., {Dahle}, H., {et~al.} 2017, \apj, 835, 5

\bibitem[{{Shen} {et~al.}(2007){Shen}, {Strauss}, {Oguri}, {Hennawi}, {Fan},
  {Richards}, {Hall}, {Gunn}, {Schneider}, {Szalay}, {Thakar}, {Vanden Berk},
  {Anderson}, {Bahcall}, {Connolly}, \& {Knapp}}]{shen07}
{Shen}, Y., {Strauss}, M.~A., {Oguri}, M., {et~al.} 2007, \aj, 133, 2222

\bibitem[{{Shen} {et~al.}(2016){Shen}, {Brandt}, {Richards}, {Denney},
  {Greene}, {Grier}, {Ho}, {Peterson}, {Petitjean}, {Schneider}, {Tao}, \&
  {Trump}}]{shen16}
{Shen}, Y., {Brandt}, W.~N., {Richards}, G.~T., {et~al.} 2016, \apj, 831, 7

\bibitem[{{Smith} {et~al.}(2009){Smith}, {Jarvis}, {Simpson}, \&
  {Mart{\'{\i}}nez-Sansigre}}]{smith09}
{Smith}, D.~J.~B., {Jarvis}, M.~J., {Simpson}, C., \&
  {Mart{\'{\i}}nez-Sansigre}, A. 2009, \mnras, 393, 309

\bibitem[{{Tapken} {et~al.}(2007){Tapken}, {Appenzeller}, {Noll}, {Richling},
  {Heidt}, {Meink{\"o}hn}, \& {Mehlert}}]{tapken2007}
{Tapken}, C., {Appenzeller}, I., {Noll}, S., {et~al.} 2007, \aap, 467, 63

\bibitem[{{U} {et~al.}(2015){U}, {Hemmati}, {Darvish}, {Mobasher}, {Nayyeri},
  {Dickinson}, {Stern}, {Spinrad}, \& {Mallery}}]{u2015}
{U}, V., {Hemmati}, S., {Darvish}, B., {et~al.} 2015, \apj, 815, 57

\bibitem[Venemans et al.(2017)]{venemans17} Venemans, 
B.~P., Walter, F., Decarli, R., et al.\ 2017, \apj, 837, 146 

\bibitem[{{Wang} {et~al.}(2006){Wang}, {Chen}, \& {Zhang}}]{wang06}
{Wang}, J.-M., {Chen}, Y.-M., \& {Zhang}, F. 2006, \apjl, 647, L17

\bibitem[{{Wang} {et~al.}(2013){Wang}, {Wagg}, {Carilli}, {Walter}, {Lentati},
  {Fan}, {Riechers}, {Bertoldi}, {Narayanan}, {Strauss}, {Cox}, {Omont},
  {Menten}, {Knudsen}, {Neri}, \& {Jiang}}]{wang13}
{Wang}, R., {Wagg}, J., {Carilli}, C.~L., {et~al.} 2013, \apj, 773, 44

\bibitem[{{York} {et~al.}(2000){York}, {Adelman}, {Anderson}, {Anderson},
  {Annis}, {Bahcall}, {Bakken}, {Barkhouser}, {Bastian}, {Berman}, {Boroski},
  {Bracker}, {Briegel}, {Briggs}, {Brinkmann}, {Brunner}, {Burles}, {Carey},
  {Carr}, {Castander}, {Chen}, {Colestock}, {Connolly}, {Crocker}, {Csabai},
  {Czarapata}, {Davis}, {Doi}, {Dombeck}, {Eisenstein}, {Ellman}, {Elms},
  {Evans}, {Fan}, {Federwitz}, {Fiscelli}, {Friedman}, {Frieman}, {Fukugita},
  {Gillespie}, {Gunn}, {Gurbani}, {de Haas}, {Haldeman}, {Harris}, {Hayes},
  {Heckman}, {Hennessy}, {Hindsley}, {Holm}, {Holmgren}, {Huang}, {Hull},
  {Husby}, {Ichikawa}, {Ichikawa}, {Ivezi{\'c}}, {Kent}, {Kim}, {Kinney},
  {Klaene}, {Kleinman}, {Kleinman}, {Knapp}, {Korienek}, {Kron}, {Kunszt},
  {Lamb}, {Lee}, {Leger}, {Limmongkol}, {Lindenmeyer}, {Long}, {Loomis},
  {Loveday}, {Lucinio}, {Lupton}, {MacKinnon}, {Mannery}, {Mantsch}, {Margon},
  {McGehee}, {McKay}, {Meiksin}, {Merelli}, {Monet}, {Munn}, {Narayanan},
  {Nash}, {Neilsen}, {Neswold}, {Newberg}, {Nichol}, {Nicinski}, {Nonino},
  {Okada}, {Okamura}, {Ostriker}, {Owen}, {Pauls}, {Peoples}, {Peterson},
  {Petravick}, {Pier}, {Pope}, {Pordes}, {Prosapio}, {Rechenmacher}, {Quinn},
  {Richards}, {Richmond}, {Rivetta}, {Rockosi}, {Ruthmansdorfer}, {Sandford},
  {Schlegel}, {Schneider}, {Sekiguchi}, {Sergey}, {Shimasaku}, {Siegmund},
  {Smee}, {Smith}, {Snedden}, {Stone}, {Stoughton}, {Strauss}, {Stubbs},
  {SubbaRao}, {Szalay}, {Szapudi}, {Szokoly}, {Thakar}, {Tremonti}, {Tucker},
  {Uomoto}, {Vanden Berk}, {Vogeley}, {Waddell}, {Wang}, {Watanabe},
  {Weinberg}, {Yanny}, \& {Yasuda}}]{york2000}
{York}, D.~G., {Adelman}, J., {Anderson}, Jr., J.~E., {et~al.} 2000, \aj, 120,
  1579

\bibitem[{{Young} {et~al.}(2014){Young}, {Eracleous}, {Shemmer}, {Netzer},
  {Gronwall}, {Lutz}, {Ciardullo}, \& {Sturm}}]{young14}
{Young}, J.~E., {Eracleous}, M., {Shemmer}, O., {et~al.} 2014, \mnras, 438, 217

\end{thebibliography}

\end{document}